
\input harvmac.tex

\def\CL{{\cal L}}
\def\calD{{\cal D}}
\def\calO{{\cal O}}

\def\zb{{\bar z}}

\def\bx{{\bf x}}
\def\bX{{\bf X}}

\def\partialb{{\bar \partial}}
\def\betab{{\bar \beta}}
\def\gammab{{\bar \gamma}}
\def\kappab{{\bar \kappa}}
\def\bb{{\bar b}}
\def\cb{{\bar c}}

\Title{UFIFT-HEP-92-26}{\vbox{
\centerline{Point-Like Interactions in String Theory}
\vskip4pt\centerline{Induced by 2-D Topological Gravity }}}


\centerline{Zongan Qiu\footnote{$^\ddagger$}{email:
qiu@ufhepa.phys.ufl.edu, qiu@ufhepa.bitnet}}

\bigskip\centerline{Department of Physics}
\centerline{University of
Florida}\centerline{ Gainesville, FL 32611 }

\vskip .7in
We consider a string theory with two types of strings with geometric
interaction.
We show that the theory contains
strings  with constant
Dirichlet boundary condition and those strings are glued together by
2-d topological gravity with macroscopic boundaries.
A light-cone string field theory is given and the
theory has interactions to all orders.

\Date{10/92}

\noblackbox

\newsec{Introduction}
It is well known that the bosonic string theory\ref\rstr{e.g. M. B. Green,
J. H. Schwarz and E. Witten, ``Superstring Theory'', Cambridge University
Press 1987.}\ is not a
consistent theory due to the existence of tachyon.
Physicists have long suspected the tachyon signals that one
considers the theory in the wrong vacuum. One of the goals of formulating
string field theory is to discover the true vacuum of string theory.
Work in this direction so far yields few insight due to our lack
of understanding of closed string field theory. Moreover string field
theory to date is formulated to generate the first quantized amplitudes,
and is therefore limited by the problems of the first quantized theory.
The question of whether there is a consistent $d>1$ bosonic string theory
is a fundamental one and deserves further exploration.
A similar question also
appears in high temperature superstring, where the Hagedorn temperature
can be thought of as the appearance of tachyon in the compactified
time\ref\raw{J. Atick and E. Witten, Nucl. Phys. {\bf B310} (1989)291.}.
Atick and Witten have further
argued that the tachyon signals a phase transition in string theory.
The tachyon problem has
been a mystery in string theory.

Recently, much progress has been made in understanding another class of string
theories, non-critical string theories of $d\le 1$\ref\rkpz{V. Knizhnik,
A. A. Polyakov and A. Zamolodchikov, Mod. Phys. Lett. {\bf A3} (1988) 819.}.
These theories are
consistent bosonic strings without a tachyon. Furthermore,
they can be understood
in term of hermitian matrix models which give a non-perturbative
definition of the theories\ref\rmatrix{
E. Br\'ezin and V. A. Kazakov, Phys. Lett. {\bf 236B} (1990) 14;
M. R. Douglas and S. H. Shenker, Nucl. Phys. {\bf B335} 91990) 635;
D. J. Gross and A. A. Migdal,
Phys. Rev. Lett. {\bf 64} (1990) 127.}.
Some of the $d\le 1$ strings are purely
topological theories, e.g. two-dimensional Topological Gravity (TG)
with $c=-2$\ref\rtop{E. Witten, Nucl Phys. {\bf B340} (1990) 281.}
\ref\rdistler{J. Distler, Nucl Phys. {\bf B342} (1990) 523.}.
Many researchers in the field suspect $d\le 1$ strings
are related in some way to critical string theory with $d=26$.
But the exact relationship of
$d\le 1$ non-critical strings with critical
string theory has
not been demonstrated.

In this paper we discuss a string theory
that incorporate non-critical
$d\le 1$ strings
into critical bosonic string theory.
In the first quantized language, the amplitude of the string theory
has new contributions from ``colored'' Riemann surfaces, black and white
in our case, which come from interactions between two types
of strings. The white region represents ordinary bosonic string
with suitable boundary condition and the black region the non-critical
strings. In fact, we will only consider
a special case where there is a different black string for each space-time
point and each one is given by
TG.
Therefore in the calculation of amplitudes in the theory one not only has to
sum over all surfaces but also has to sum over the coloration
and all possible black strings as well.

We will describe the gauge fixed theory in this paper and
postpone the gauge invariant formulation to later publication.
Nevertheless, some formal properties of the gauge invariant theory
will be addressed. The gauge fixed theory is a string theory containing strings
with constant Dirichlet Boundary Condition(DBC).
Some general properties of the theory with DBC was studied by
Green and Shapiro\ref\rgrsp{M.B. Green and J.A. Shapiro, Phys. Lett.
{\bf 64B} (1976) 454; M.B. Green, Phys. Lett. {\bf 69B} (1977) 89.}.
But in our theory
the point like object comes from TG.
The interactions are given by string
geometrical interactions and determined to all orders.

The organization of the paper is as follows. In section 2. we
will describe the first quantized theory and review the
relevant results
of the TG. In section 3,
the main ingredients needed to describe our theory are discussed.
A string field theory in light-cone gauge is given in section 4.
Various issues related to our results are addressed in section 5.

\newsec{The Configuration Space of String Theory and 2-D Topological Gravity}

The Polyakov description of string theory starts with the function integral
\eqn\epoly{
Z = \int \calD g \calD X e^{-S[g,X]},
}
with the Euclidean action
\eqn\epolyact{
S[g,X] = {1\over 8\pi} \int d^2\xi \sqrt{g} \left(g^{ab}\partial_a X^\mu
\partial_b X_\mu - 4 {\mu_0}\right),
}
where $X_\mu(\xi)$ is the location of the surface in space-time.

Before we proceed further, we would like to argue that the function integral
\epoly\
contains contribution where the whole world sheet is mapped to a single
space-time point, i.e. constant map.  To show that, we insert a product of
$\delta$-functions
$\prod_{\xi} \delta(X^\mu(\xi)-x^\mu)$ into the path integral, which singles
out a constant map in the configuration space.
The partition function
reduces to that of two dimensional gravity with no matter, i.e.
pure 2-d gravity theory
\eqn\epure{
Z^{top}_x = \int \calD g \calD X e^{-S[g,X]}
\prod_{\xi} \delta(X^\mu(\xi)-x^\mu) = \int \calD g e^{-S'[g]}
}
where the action $S'$ is given by \epolyact\ without $X$ dependent terms.

In fact, the pure two dimensional gravity is described by a TG
which can be thought of as $c=-2$ matter coupled with 2-d
gravity \rdistler\ref\rtop2{E. Verlinde and H. Verlinde, Nucl. Phys. {\bf B348}
(1991) 457.}.
Therefore the configuration space of \epoly\ contains configurations
corresponding to one TG for every space-time
point $x^\mu$.

More generally,
the configuration space contains configurations which map
regions, which will be denoted black regions,
of world-sheet to single points in space-time and rest of world-sheet,
denoted as white regions,
to a surface in space-time. The region of the world-sheet mapped
to single space-time point is described by TG
with a suitable boundary condition.
Therefore we divide the configuration space into three parts:

i). constant maps, topological sector.

ii). generic maps, string sector.

iii). mixed maps, interacting sector.

We will assume that all three sectors are important in our theory
and their relative weights are governed by a yet undetermined coupling constant
$\lambda$ which we will discuss in detail later.

String sector
of the configuration space was considered solely in the usual
formulation of string theory
and one implicitly assumed the other two
sectors had zero measure and therefore could be ignored. One of the reasons for
such an assumption is
that the string with single point as target space is trivial. Recent
advances in $d\le 1$ string theory show that this is not the case\rkpz.
Therefore it
is necessary to re-examine this question. Another possibility is that
the non-perturbative effects in string theory amplify the contribution of
these configurations and one can therefore put them in from beginning
in perturbation theory. It is also conceivable that the theory is a total
different string theory which is also certainly interesting.

The descriptions of i). and ii). are straightforward and therefore we
will concentrate on iii). In string theory,
the interesting observables are vertex operators integrated over the surface
of the world-sheet. For these observables, the modifications in our
theory come from contributions from the interacting sector.
A general new contribution comes from a string diagram of
colored surface Fig.1,
and the path integral is performed in the the following way.
For the white regions on the surface we integrate over $X$
with DBC and $g$ with proper gauge fixing.
For the black regions we integrate only over $g$ with
action $S'$ in the proper gauge, and then integrate over the location $x$.
The vertex operators will only be integrated over the
white regions of the surface.
The path integral can
also be computed by inserting a product of $\delta$-functions
for every point of the black region in \epoly, and then integrating over the
different boundary values $x$ for each boundary as
$$
A(y_1, y_2, \cdots) = \int \calD X\calD g \prod_i dx
e^{-S} \prod_j \delta(X(\xi_j) - y_j) \prod_{\xi\in \Sigma_i}
\delta(X(\xi)-x_i)
$$
where $\Sigma_i$' are black regions of world-sheet.
Because black regions are point-like
in space-time, the interactions are local in space-time.

One can also view the black and white regions to represent two different
types of strings, will be referred as string and topological string.
In doing so, a colored surface represents interactions
of strings and topological strings. This is analogy to the situation of
interaction of different particles
in first quantized theory. Therefore we associate
a coupling constant $\lambda$ for each boundary between the two regions.
This is the first example of a string theory with two types of
strings interacting locally in space-time.
It is also important to point out the interaction is not local on
the world-sheet, which is certainly not a requirement in string theory.

Our main goal now is to develop a formalism to study the above theory
in a systematic way. It is convenient to separate the problem into
three parts. The white region(string part) is the well
known object in string theory with suitable boundary condition.
The black region(topological string part) is a TG with suitable boundary
condition and the proper boundary conditions to glue
those regions together.

What is the proper boundary condition?
The answer actually depends on the particular gauge
used. In conformal gauge,
to maintain the BRST invariance one should require the
matter stress-tensor to be continuous cross the boundary. On
the string side, because the whole boundary is being mapped to
a single point, the natural boundary condition is
the DBC for $X^\mu$. The ghost fields are shared on both sides, and
therefore they are continuous across the boundary.
The correct procedure to implement
these boundary conditions requires the introduction of auxiliary fields on the
string part of the world sheet and will be addressed in a
subsequent publication.
We will see later that
the conformal gauge is not the most efficient gauge to use
in our theory for practical calculations. However, it is a good gauge to
use to study the formal properties of the theory.

The main results of TG relevant for our purposes can be summarized
as following\rdistler.
In the conformal field theory description of TG, one starts with spin-2
bosonic ghost system together with the fermionic ghost with action
\eqn\ebcbg{
S = S_{bc} + S_{\beta\gamma} = {1\over\pi} \int b\partialb c + \bb\partial
\cb + \beta \partialb\gamma + \betab\partial\gammab
}
It is more convenient to bosonize the $\beta-\gamma$ ghost by
\eqn\eboson{
\beta = e^{-\phi(z)} \partial \xi,\qquad \gamma = e^{\phi(z)} \eta
}
following Friedan, Martinec and Shenker\ref\rfms{D. Friedan, E. Martinec and
S. Shenker, Nucl.Phys. {\bf B271} (1986) 93.}.
The cosmological constant term is  expressed in term of ``Liouville'' field
$\phi$ as
\eqn\ecosmo{
S_{cosmo} = -{1\over\pi} \int \sqrt{\hat g} e^{-\phi(z,\zb)}
}

One can further bosonize $\eta-\xi$ system, $\eta = e^{i\psi}, \xi =
e^{-i\psi}$
and the dressed primary fields which will be our observables are
\eqn\eobles{
O_n = e^{in\psi(z,\zb)} e^{(n-1) \phi(z,\zb)}.
}
The correlation functions of these operators are not the main focal point
here. The corrections to the amplitude of these operators
due to the interacting
sector can also be addressed by the formalism discussed below.
Of course, in that case, we should sum over the contributions of the white
regions instead.

\newsec{Modification of String Theory}

It turns out the simplest gauge to work is the GGRT\ref\rggrt{J. Goldstone,
P. Goddard, C. Rebbi and C. Thorn, Nucl Phys.
{\bf B56} (1973) 205.}\ light-cone gauge.
The first quantized string is formulated in terms of
coordinates $X^\mu(\sigma, t)$ satisfying two dimensional wave equation
\eqn\ewave{
\left( {\partial^2\over \partial t^2} -{\partial^2\over
\partial \sigma^2}\right)
X^\mu(\sigma,t) = 0}
which sweep out a two dimensional surface.
The reparametrization invariance of the world sheet implies the phase
space of the string is subject to the first class constraints
\eqnn\ect{
$$\eqalignno{
{\dot X}^2 + X'^2 &= 0,\cr
{\dot X}\cdot X' &= 0.&\ect\cr}
$$
}
In the light-cone gauge a relativistic string
$X^\mu(\sigma)$ can be represented in terms of only its transverse components,
and hence the question of ghost never appears. Their constraints
imply $X^+ = {1\over \sqrt 2} (X^0 + X^{D-1}) = \tau$ independent of $\sigma$.
All coordinates are determined in terms of physical operators
$\bX(\sigma)$, ``transverse'' components of $X^\mu$, i.e. $\mu = 1,2,\cdots,
D-2$ and thereby eliminating the redundant longitudinal modes.
In the light-cone gauge, the matter stress-energy tensors are set
to zero by the constraint condition \ect. Another advantage of the light-cone
gauge is that the only consistent closed string field
theory is formulated in this gauge.

On the normal string part, the
boundary condition is given by $\prod_s\delta(\bX(s) - \bx)$,
and the value of $X^+$ along the boundary.
On the topological string part, we will work in the conformal
gauge.  The fields $\beta-\gamma$ and $b-c$ of the TG are restricted to
the black region of the surface and therefore the
correct boundary condition is the Neumann boundary
condition for both $\beta-\gamma$ and $b-c$. That is the boundary
condition we are going to use for the black region.
We will refer this gauge choice as ``mixed gauge''.

It is assumed in critical bosonic string theory that
the Liouville field decouples from the theory due to Weyl invariance.
The Liouville action is multiplied by $D-26$, which vanishes
in the critical dimension.
Therefore it is clear in our case that the boundary condition for the
Liouville field of the TG is the
Neumann boundary condition. The boundary condition of the Liouville
field will be very different
if the white region of the theory is not in the critical dimension.

In the mixed gauge, the contribution of many string diagrams
with the black region that have the same number of boundaries but different
genus, moduli, etc. can be computed together by using the matrix model
method\ref\rmtbd{J. D. Edwards and I. R. Klebanov,
Mod. Phys. Lett. {\bf A6} (1991) 2901; G. Moore Nucl. Phys. {\bf B377}
 (1992) 143.}.
The main advantage
of the matrix model method is that one can evaluation the integral
$$
W^{(n)} \sim \sum_{\Sigma^n_g} \int \calD g e^{-S'}
$$
directly, where the sum is over all possible surfaces with $n$ boundaries.
First we compute
the sum over all surfaces with $n$ macroscopic boundaries
of fixed lengths in TG. It is given by\rmtbd
\eqnn\emacro{
$$
\eqalignno{
W^{(n)}_l |ln \Delta|&= (-1)^{n+1} |ln \Delta| 2^{3-n} \mu^{n/2}
(-96 \kappab^2)^{-1+{n\over 2}}\cr
&\times \sum_{g=0}^\infty {\kappab^{2g}\over g!} (l\sqrt{\mu})^{3g-3+n}
I_{3g-3+n}(l\sqrt{\mu}),&\emacro\cr
}$$
}
where $l=\sum l_i$ is the total length of the boundaries.
$\mu$ the renormalized cosmological constant and $\kappab^2 = \kappa^2/\mu^3$
is the dimensionless loop counting parameter.
$\kappa$ is the the renormalized string coupling constant of TG and $I$ are
the modified Bessel functions.

Notice that $W^{(n)}_l$ depends only on the total length
of the boundaries.
Summation over all possible boundary lengths can be easily
done by introducing a boundary cosmological constant $\xi$
\eqn\ew{
W^{(n)} = \int dl_1 \cdots dl_n e^{-\xi l} W^{(n)}_l
}
$W^{(n)}$ is a function of $\mu,~\xi,~\kappab^2$
and contains contribution of all possible surface with $n$ boundaries.

For each boundary of black surface we glue to it a boundary of
white surface. The incredibly
huge amount of string diagrams are greatly simplified.
In space-time the contributions of topological string correspond to
introducing new point-like interactions in string theory.
Unfortunately there
are still a very large number of string diagrams to be considered.
One also expects that the interesting phenomena come from a partial
(infinite) summation over
contributions from these diagrams. The first quantized formalism is not
very good at handling this question efficiently. In the next section we
will discussion how to incorporate these diagrams in the second quantized
string field theory via new string interactions.

\newsec{Light-Cone String Field Theory}

We start by summarizing the main results of the light-cone
bosonic string field theory.
In this gauge, the string field functional $\Psi[X(\sigma)]$
is independent
of the longitudinal components except the zero modes of $X^+, X^-$.
One then takes the Fourier transform with respect to the zero mode of $X^-$,
so the field depends on $P^+$, the width of the string. The string field
functional becomes $\Psi_{P^+} [\bX, \tau]$.
The region of the parameterization of the string is $[0, \pi P^+]$.
The Lagrangian for the closed bosonic string field theory\ref\rthorn{
C. B. Thorn, Nucl. Phys. {\bf B263} (1986) 493.}\ follows simply from that of
light-cone open string field theory
\rstr\ref\rlcst{
M. Kaku and K. Kikkawa, Phys. Rev. {\bf D10} (1974) 1110;
M. Kaku and K. Kikkawa, Phys. Rev. {\bf D10} (1974) 1823.
E. Cremmer and J. L. Gervais, Nucl. Phys. {\bf B90} (1975) 410.}\
to reproduce the results of
the Mandelstam's interacting string formalism\ref\rmand{
S. Mandelstam, Nucl. Phys. {\bf B64} (1973) 205; {\bf B69}
(1974) 77; {\bf B69} (1974) 109.}.
The Lagrangian is given by
\eqn\elag{
\CL = \CL_0 +\CL_1 = \Psi^\dagger K \Psi
+ \half g ( \Psi^\dagger\circ\Psi\circ\Psi +
h.c.)
}
in the operator formalism. The quadratic term will reproduce
the first-quantized Schr\"odinger equation
\eqn\esch{
i {\partial \Psi_{P^+} \over \partial \tau} [\bX, \tau]=
{\pi \over 2} \int_0^{\pi P^+} \left[
-{\delta^2\over \delta \bX(\sigma) ^2} + \left({\bX'(\sigma)\over \pi}\right)^2
\right] \Psi_{P^+} [\bX, \tau] \equiv h \Psi_{P^+} [\bX, \tau]
}
The $\circ$ operation in three string interaction
Lagrangian $\CL_1$ is defined by the geometrical
three string interaction vertex.

The modification of the string field theory from the point-like interactions,
which describe the contributions of the topological strings,
can be accomplished by adding
new interaction terms, in fact infinitely many of them, to the
bosonic string Lagrangian. On the quadratic level, there are
two new interacting
terms that are obtained
by looking at the first quantized amplitude with two vertex operators
inserted. They are
\eqn\est{
\CL^{(1)}_2 + \CL^{(2)}_2 = \Psi^\dagger \calO_1 \Psi + \Psi^\dagger \calO_2
\Psi
}
as illustrated in Fig.2. $\CL^{(1)}_2$ represents an interaction where
at given $X^+$ a segment of string carrying finite $P^+$ becomes transversely
localized, i.e. $\bX(\sigma) = \bx$ along the segment.
It has been shown by Green and Shapiro\rgrsp\ that such an interaction is
point-like in space-time and therefore Lorentz invariant. $\CL^{(1)}_2$
is also proportional to $W^{(1)}$ to take into account the contributions
of all black regions with one boundary. We also introduce a new coupling
constant $\lambda$, and its power counts the number of boundaries for the
interaction. Of course, we have to sum over all possible such interactions
by integrating over the length  and the locations of the segment on the string.

It is important to point out that we also have to integrate over the locations
of the interaction over space-time. In doing so, we can ignore the $ln|\Delta|$
by interpreting it as the density of topological strings
in space-time.
In the matrix model, $\Delta$ is introduced as energy gap which is
taken to zero in the double scaling limit. Similar treatment will be done for
other interactions.

$\CL^{(2)}_2$ is the interaction where the whole string becomes transversely
localized at some instant $X^+$. Summing over the contributions from
all black surfaces with two boundaries and putting in the proper power of
coupling constant $\lambda$, it is given explicitly by
\eqnn\elc{
$$
\eqalignno{
\CL^{(2)}_2 &= \half \lambda^2 W^{(2)} \int\prod_{i=1}^2
{d P^{+_i}}
\calD \bX_i(\sigma) d\bx \delta(P^+_2-P^+_1)
\Psi^\dagger_{P^+_2}[\bX_2(\sigma),\tau]
\Psi_{P^+_1}[\bX_1(\sigma), \tau]\cr
&\times\prod_{\sigma} \delta(\bX_2(\sigma) -
\bx)\delta(\bX_1(\sigma) - \bx) + h.c &\elc\cr
}$$
}

Few remarks are in order before we go on to describe higher order interactions.
The interaction $\CL^{(1)}_2$ was considered by Green and Shapiro
\rgrsp, and later Green\ref\rgreen{
M. B. Green, Nucl. Phys. {\bf B116} (1976) 449; {\bf B124} (1977) 461.}\ as
the emission of zero momentum scalar particles into vacuum.
They showed that the interaction is truly
point-like in space-time and that it can modify the
large-angle elastic scattering
in the open string when considering the leading contribution to the scattering
amplitude.
In their original formalism, $\CL^{(2)}_2$ is negligible
because it is the limiting case of $\CL^{(1)}_2$ when the whole string
becomes transversely localized.

In our formulation, the interactions are
consequences of normal strings interacting with
topological strings. Because the interactions are geometric, the consistent
condition should be much easier to check. In particular $\CL^{(2)}_2$
takes into account infinitely many surfaces with two boundaries which are
different surfaces from those
of $\CL^{(1)}_2$, and therefore must be included. In fact, the first quantized
picture tells us exactly what the new interaction terms are in a given
order of string field theory.

There are new cubic interaction terms in our theory,
$\CL^{(2)}_3 + \CL^{(3)}_3$,
which are represented in Fig.3. The four-string interacting terms
are $\CL^{(2)}_4 + \CL^{(3)}_4 + \CL^{(4)}_4$, as shown in Fig. 4.

It is very important to note that for every boundary
there are at most two light-cone strings simultaneous touching it.
This limits the number of interactions a black region with $n$ boundaries
can generate, e.g. there are new interaction terms proportional
to $W^{(n)}$ which couple from $n$ to $2n$ strings.
In particular there are only a finite number of
new diagrams for $N$ string tree amplitude
when we use the correct propagator $(K+\calO_1 + \calO_2)^{-1}$ from the
quadratic part of the theory.

The propagator $(K+\calO_1 + \calO_2)^{-1}$ summarizes the contributions from
infinitely many string diagrams. Because $W^{(n)}\propto (-1)^{n+1}$, the
effect of $\calO_1$ and $\calO_2$ could shift the masses square upward
for real coupling constant $\lambda$. The determination of the mass
spectrum of our theory will be addressed in a future publication.

\newsec{Conclusions and Remarks}

One of the obvious questions is whether the new terms included in string field
theory preserve the unitarity of the string amplitudes. The interaction of
strings with point-like structure has been studied\rgrsp\rgreen.
The major difference is that we give an explicit realization of the
point-like structure, i.e. induced by TG.
Therefore the properties of the point-like structure are determined.
The point-like structure couples with a segment of string and the coupling
is geometric in the two dimensional world-sheet. Nevertheless
Green's\rgreen\ discussion about the unitarity of scattering amplitude still
applied. Our theory is at least perturbative unitarity.

What is the fundamental reason for the existence of the mixed sector
of the string theory? The following scheme is one possible answer to
this question.
As we mentioned early, the configuration space divided into three sectors.
At very high energy, or high temperature, the whole theory
is given by the topological sector. The theory consists of product
of TG for each space-time point.
For each TG, there is a BRST symmetry(string symmetry) and the symmetry
of the theory is the product
of these BRST symmetries.
Moreover, the theory is topological in space-time
and has no continuous degree of freedom in
the normal sense.
At lower energy the physics is given by the theory we described in the main
text and the symmetry, when we work in conformal gauge, is one BRST symmetry.
The natural question to ask is where are all
the continuous degrees of freedom come from? We expect the partial answer
is that they coming from symmetry breaking. The exact dynamical question
of how this process occurs is not clear.
The theory we describe is only an effective
theory after symmetry breaking.

The extension of our results to superstring is of considerable interest.
Instead of topological gravity, we should consider topological
supergravity
as the topological sector of superstring. By including the topological sector
and interacting sector to superstring, it is conceivable that
the theory automatically gives an expectation
value to dilaton and breaks supersymmetry.
The reason is that the point-like structure from the topological
string couples only with scalar fields of
superstring.  Clearly, more work in this direction is needed to clarify
this issue.

The mixed sector is also playing a crucial role when considering the
non-critical string theory of $d>1$. In this case the light-cone
gauge for the string sector is not well understood and the gauge to
work in is the conformal gauge. The boundary condition of the Liouville
field is the central issue.
When we use the
prescription of the
Liouville field in terms of a free field, with suitable coupling
to the curvature\ref\rlioufree{T.L. Curtright
and C.B. Thorn, Phys. Rev. Lett. {\bf 48} (1982) 1309.},
the boundary condition is simply the continuation
condition of the Liouville field. The scaling behavior of the
partition function, $Z(\mu) \sim \mu^{2-\gamma}$,
can be easily calculated\ref\rqiu{ Z. Qiu, to appear.}\ as
in \ref\rddk{F. David Mod.
Phys. Lett. {\bf A3} (1988) 1651;
J. Distler and H. Kawai, Nucl. Phys. {\bf B321} (1988) 509.}\ with
\eqn\esus{
\gamma = 2 - \sqrt{25-d\over 3}
}
for the spherical topology.

It is reasonable to expect that
the topological sector of the theory describes the theory at higher
temperature. If we take the point of view that the
the statistical mechanics of string theory at finite temperature
is described also by Euclidean string theory with the time direction
compactified on a circle\raw, then at very high temperatures, above
Kosterlitz-Thouless (KT) transition temperature,
the theory is a collection of TG theories. When we go below the KT
transition temperature, the high temperature string theory will be
described by collection of one dimensional (the ``time'' direction) strings
for every space point. In this case the free energy per unit
volume can be computed by using the matrix model technique.
It is given\ref\rgk{D. J. Gross and I. R. Klebanov, Nucl. Phys. {\bf B344}
(1990) 475.
},
at least when the the temperature is low enough, by
\eqn\efree{
{f\over T} \sim {1\over 4 \pi}
{1\over T g_{st}^2}  - {1\over 36\pi } \left(\pi^2 T
+ {1\over T}\right) + \cdots,
}
which can be written in a more suggestive form when we use the
effective string coupling constant $g_{eff} = T g_{st}$ as
\eqn\efreeb{
{f\over T} \sim {1\over 4 \pi}
{T\over g_{eff}^2}  - {1\over 36\pi } \left(\pi^2 T
+ {1\over T}\right) + \cdots,
}
where we absorb the $ln \mu$ into the definition of volume as before.
It has precisely the same form as
suggested by Atick and Witten\raw, from a string calculation.
The dominant contribution is from genus zero.
It therefore offers another piece of evidence that the high
energy phase of strings is described by topological string theory.

\bigbreak\bigskip\bigskip\centerline{{\bf Acknowledgments}}\nobreak

I would like to thank
P. Griffin, M. McGuigan S. Sin, and especially C. Thorn for
valuable discussions.
This work was supported in part by the United States Department of
Energy under contract No. FG05-86-ER40272.

\bigskip

\listrefs

\bye